%% file: make_astro.tex
\documentclass{mpe_report}

\usepackage{psfig,graphicx,epsfig}
\usepackage{color}
\usepackage{amsmath,amssymb,epic,eepic,array}

\begin{document}

\pagenumbering{arabic}
\setcounter{page}{149}

 \renewcommand{\FirstPageOfPaper }{149}\renewcommand{\LastPageOfPaper }{152}\include{./mpe_report_bucciantini}              \clearpage

\end{document}

%% file: mpe_report_bucciantini.tex
\title{Relativistic MHD Winds from Rotating Neutron Stars}
\author{N.Bucciantini\inst{1}, T.A. Thompson\inst{2}, J.Arons\inst{1}, E.Quataert\inst{1}}  
\institute{Astronomy Department, University of California at Berkeley, 
601 Campbell Hall, Berkeley, CA, 94720-3411, USA
\and  Department of Astrophysical Sciences, Peyton Hall, Ivy Lane, Princeton, NJ, 08544, USA}
\maketitle

\begin{abstract}
 We solve the time-dependent dynamics of axisymmetric, general relativistic MHD
 winds from rotating neutron stars. The mass loss rate is obtained self 
consistently as a solution of the MHD equations, subject to a finite thermal
 pressure at the stellar surface. Conditions are chosen to be representative 
of the neutrino driven phase in newly born magnetars, which have been considered
 as a possible engine for GRBs. We compute the angular momentum and energy 
losses as a function of $\sigma$ and compare them with the analytic expectation 
from the classical theory of pulsar winds. We observe the convergence to the 
force-free limit in the energy loss and we study the evolution of the closed
 zone for increasing magnetization. Results also show that the dipolar magnetic
 field and the presence of a closed zone do not  modify significantly the 
acceleration and collimation properties of the wind. 
\end{abstract}

\section{Introduction}

Magnetically dominated outflows from stars and accretion discs, are a key element
 in understanding the  evolution and properties of these objects. Magnetic
 outflows are ubiquitous in powering a variety of astrophysical systems, by 
converting the rotational energy into wind kinetic energy. Rotationally powered
 pulsars are among the best example of relativistic outflows produced by a fast 
compact rotator. The high magnetic field at the surface ($B\sim 10^{10-13}$ G) and
 short rotation period ($P\sim 0.001-1$ s) cause particles injection.
 If the charge density  in the magnetosphere of the neutron star, due to pair 
creation, exceeds the Goldreich-Julian value then conditions can be modeled in 
terms of ideal RMHD.

Observations of the interaction of the pulsar wind with the surrounding SNR,
 tell us that far away from the source the wind is ultrarelativisitic 
($\gamma\sim 10^{4-6}$) and weakly magnetized (\cite{arons04}),
 in contrast with conditions inside the Light Cylinder, where $\gamma\sim 100$, 
and magnetization is high. This implies that the high Lorentz factor is achieved 
by almost completely converting magnetic energy into kinetic energy.

In this sense, pulsars constitute the prototype of relativistic accelerators:
 a plasma is assumed at the base with small outflow velocity (compared to the 
asymptotic velocity), embedded in a strong magnetic field. Magnetorotation then 
accelerates the plasma and converts the magnetic energy into relativistic motion. 

In the case of pulsars, the plasma density at the surface of the neutron star, 
drops to very low values, and the magnetosphere is strongly magnetically dominated.
 Such conditions are close to force-free, and the equations that describe the 
flow dynamics become stiff, thus making impossible for present day codes to 
model such environment, in terms of MHD. Due to the presence of gaps at the 
foot points of open field lines, plasma is injected in the magnetosphere at a
 supersonic speed, quite differently from what happen in normal stars, where 
injection is subsonic.   

\begin{figure*}  
\centerline{\psfig{file=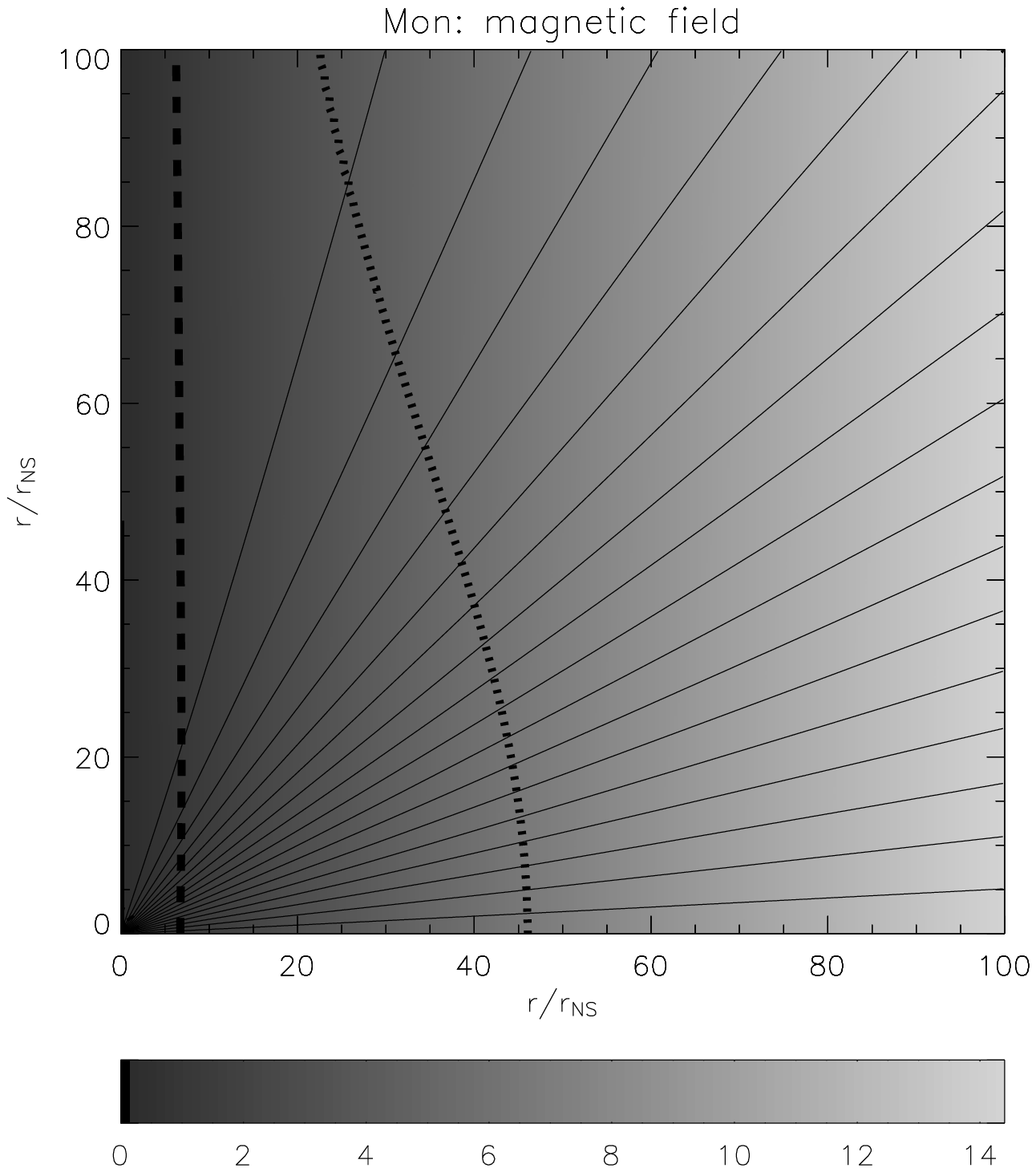,width=5cm} 
\psfig{file=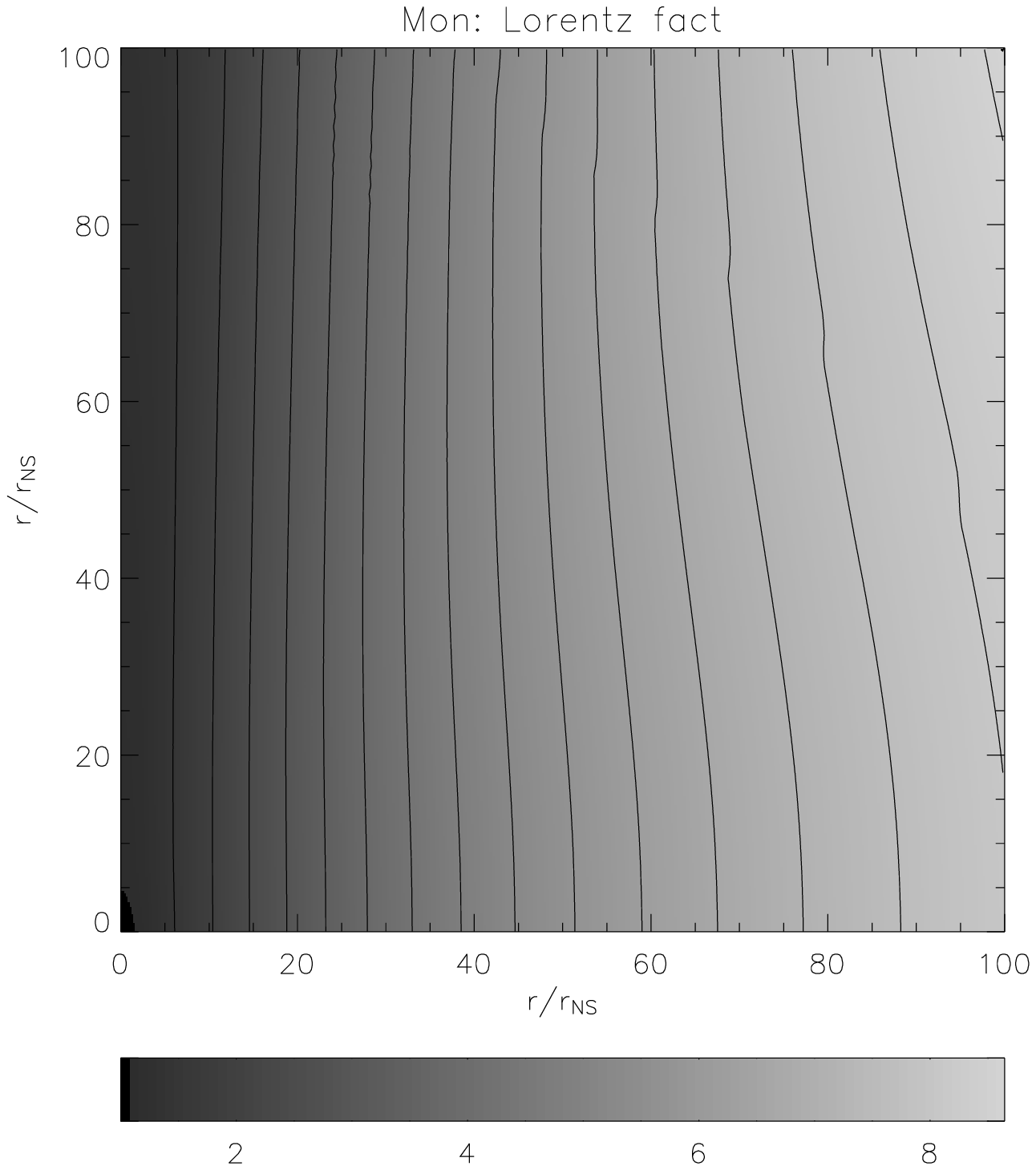,width=5cm} 
\psfig{file=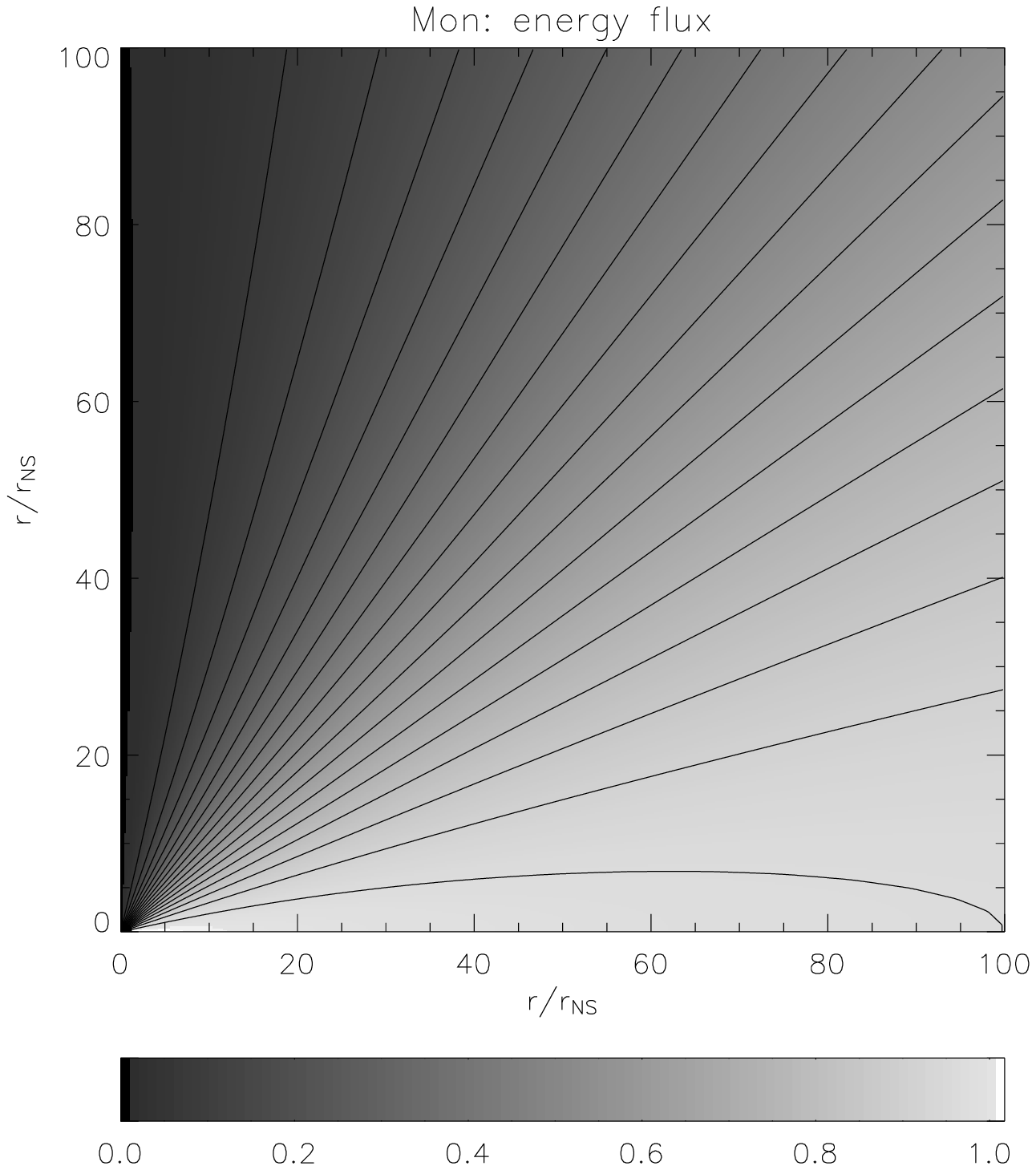,width=5cm}   }
\caption{Wind structure for a monopolar magnetic field, $\sigma=200$, $\Omega=0.14$. 
 Left: poloidal magnetic field lines, and ratio $|B_{\phi}/B_r|$;  
 Alfvenic (dashed) and fast (dotted) surfaces. Center: Lorentz factor, $\sim\sin{(\theta)}$. 
 Right: energy flux in normalized units, $\sim\sin^2{(\theta)}$. From \cite{buc06}.}
\end{figure*}

However low magnetized neutron star winds are of interest
primarily in understanding the physics of very young neutron stars. In
the seconds after collapse and explosion, the neutron star is hot
(surface thermal speed  $\sim 0.1-0.03 c$ ), and has an ``extended atmosphere''.
  This proto-neutron star cools and contracts on its Kelvin-Helmholtz
timescale ($\tau_{\rm KH}\sim10-100$\,s), radiating its gravitational
binding energy ($\sim$$10^{53}$ ergs) in neutrinos of all species
(\cite{pons99}).  The cooling epoch is
accompanied by a thermal wind, driven by neutrino energy deposition,
 which emerges into the post-supernova-shock ejecta (\cite{thom01}).

After birth, the thermal pressure at the edge of the
protoneutron star surface  decreases sharply
as the neutrino luminosity decreases on a timescale
$\tau_{\rm KH}$, as the star cools and deleptonizes.   
At some point during the cooling epoch the magnetic energy density
will exceed the thermal pressure, and the wind region will
become increasingly magnetically dominated. For
magnetar-strength surface fields ($B_0\sim10^{14}-10^{15}$ G) the
magnetic field dominates the wind dynamics from just a few seconds
after the supernova (\cite{thom03}).

If magnetars are  born with millisecond rotation periods
(\cite{thom93}), their rotational energy can be very large
($\sim$$2\times10^{52}$ ergs) relative to the energy of the supernova
explosion ($\sim$$10^{51}$ ergs) and because the spindown timescale can be
short for larger value of the Alfvenic radius, these proto-magnetar 
winds have been considered
as a mechanism for producing hyper-energetic supernovae 
(\cite{thom04}). Because the wind becomes increasingly
magnetically-dominated, the outflow must eventually become
relativistic. For this reason, proto-magnetar winds have also been
considered as a possible central engine for long-duration GRBs (e.g., \cite{thom04}).

In the simple 1D radial model for relativistic winds from compact rotators 
by \cite{michel69}, solutions can be parameterized in terms of 
$\sigma=\Omega^2\Phi^2/\dot{M}$, 
where $\Omega$ is the rotation rate, $\Phi$ the magnetic flux and $\dot{M}$ 
the mass flux. $\sigma$ is the maximum Lorentz factor the wind can achieve if 
all magnetic energy is converted into kinetic energy, and at the base of a pulsar
 it is $\sim 10^6$. The results by \cite{michel69} show that the asymptotic 
Lorentz factor tends to $\sigma^{1/3}$. This is known as the $\sigma-\gamma$ 
problem (\cite{arons04}). If the flow is allowed to diverge more than radially  
then acceleration is more efficient (\cite{beg94}). The presence of a 
magnetic field can cause collimation of the flow along the rotation axis, 
and divergence along the equator, so in principle one might expect magnetic
 collimation to be also responsible for the acceleration of the wind. However
 when the flow speed reaches values close to $c$ the Coulomb term in the 
electromagnetic force cannot be neglected and it balances the magnetic hoop
 stress. Solutions of the relativistic Grad-Shafranov equation for a 
monopolar field in the trans-fast regime (\cite{beskin98}) have shown that 
acceleration is too slow and the asymptotic results of the Michel model 
apply also in 2D axisymmetric geometry, meaning that the flow remains radial. 

\begin{figure*}  
\centerline{\psfig{file=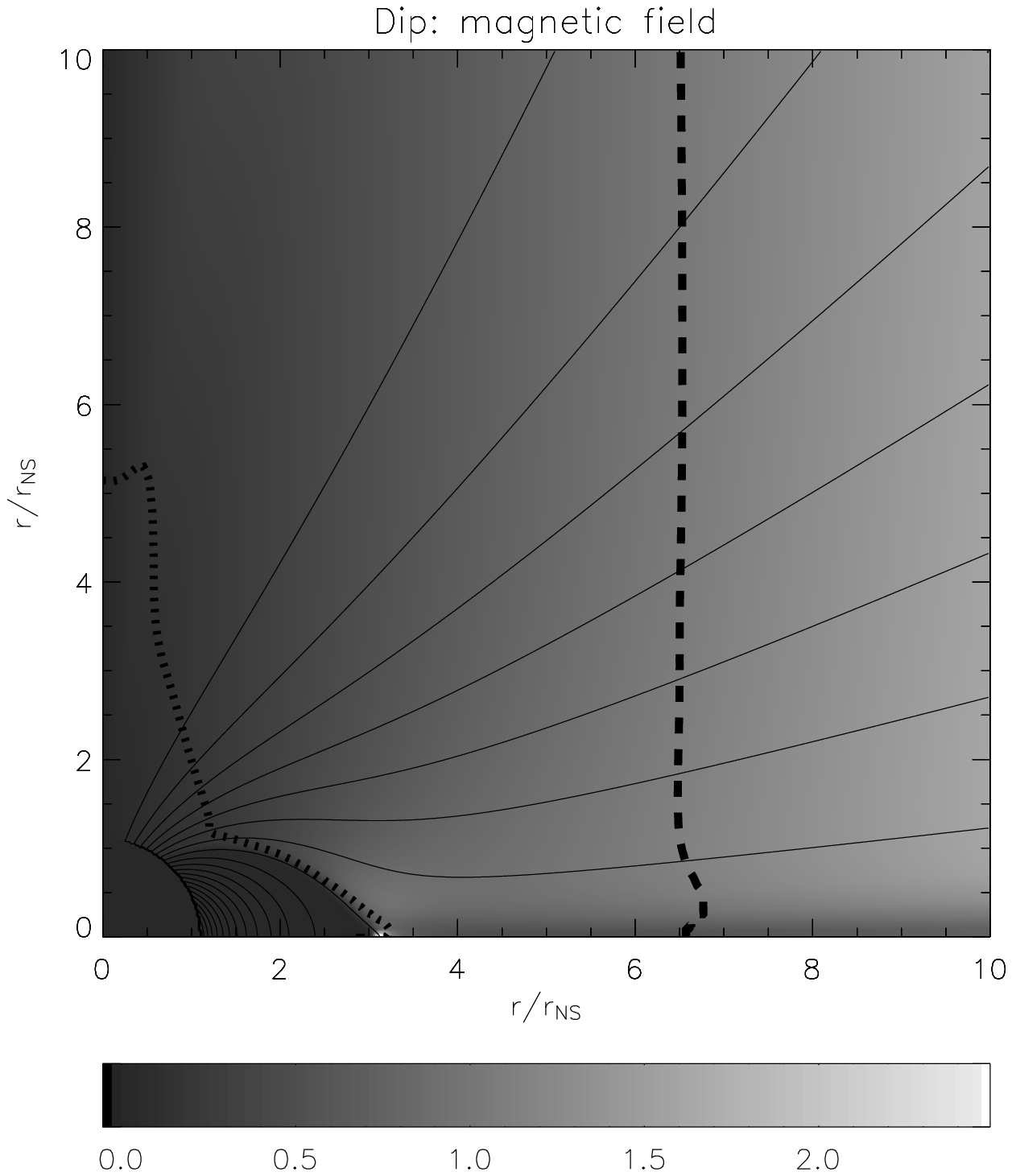,width=5cm} 
 \psfig{file=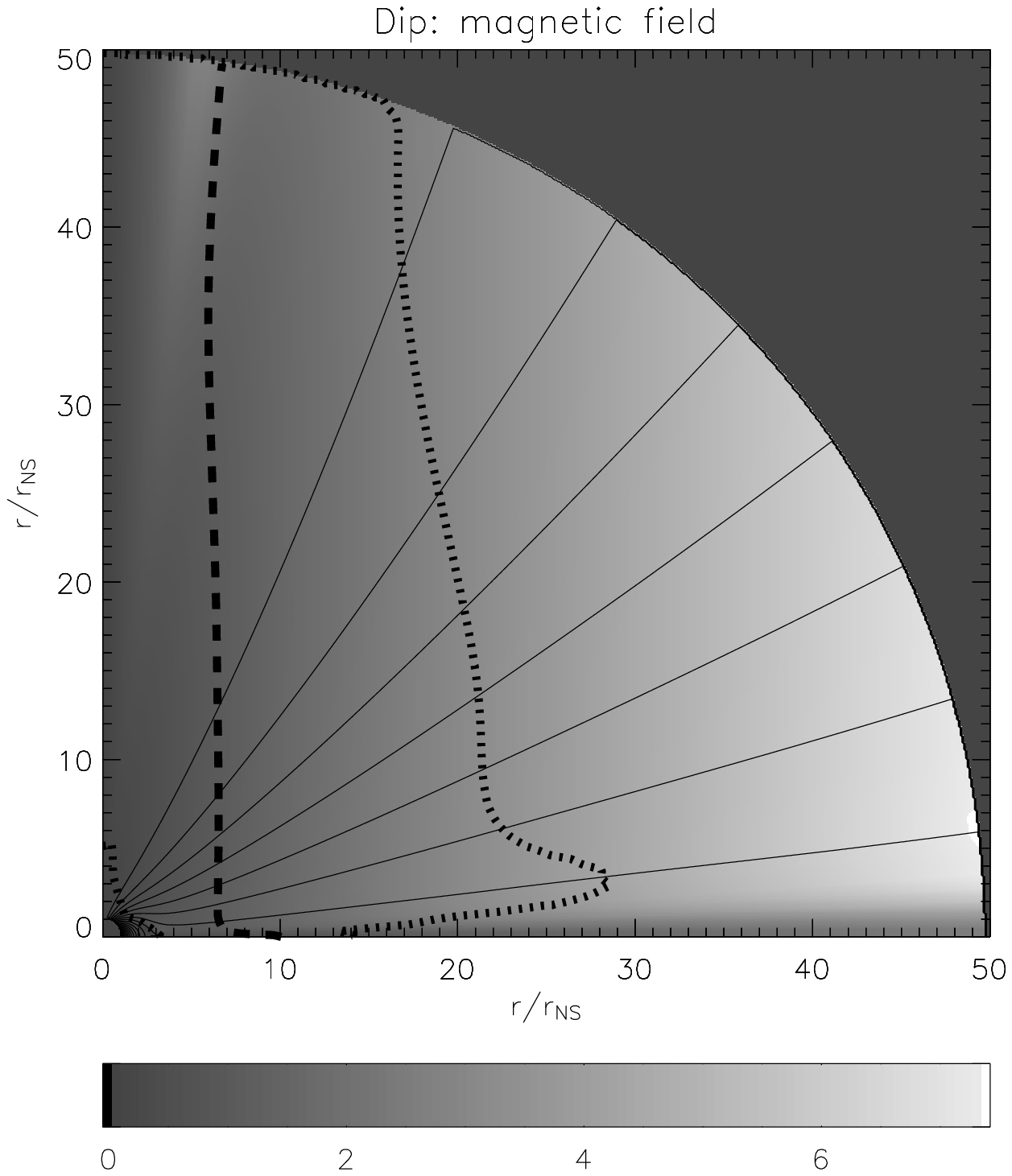,width=5cm} 
 \psfig{file=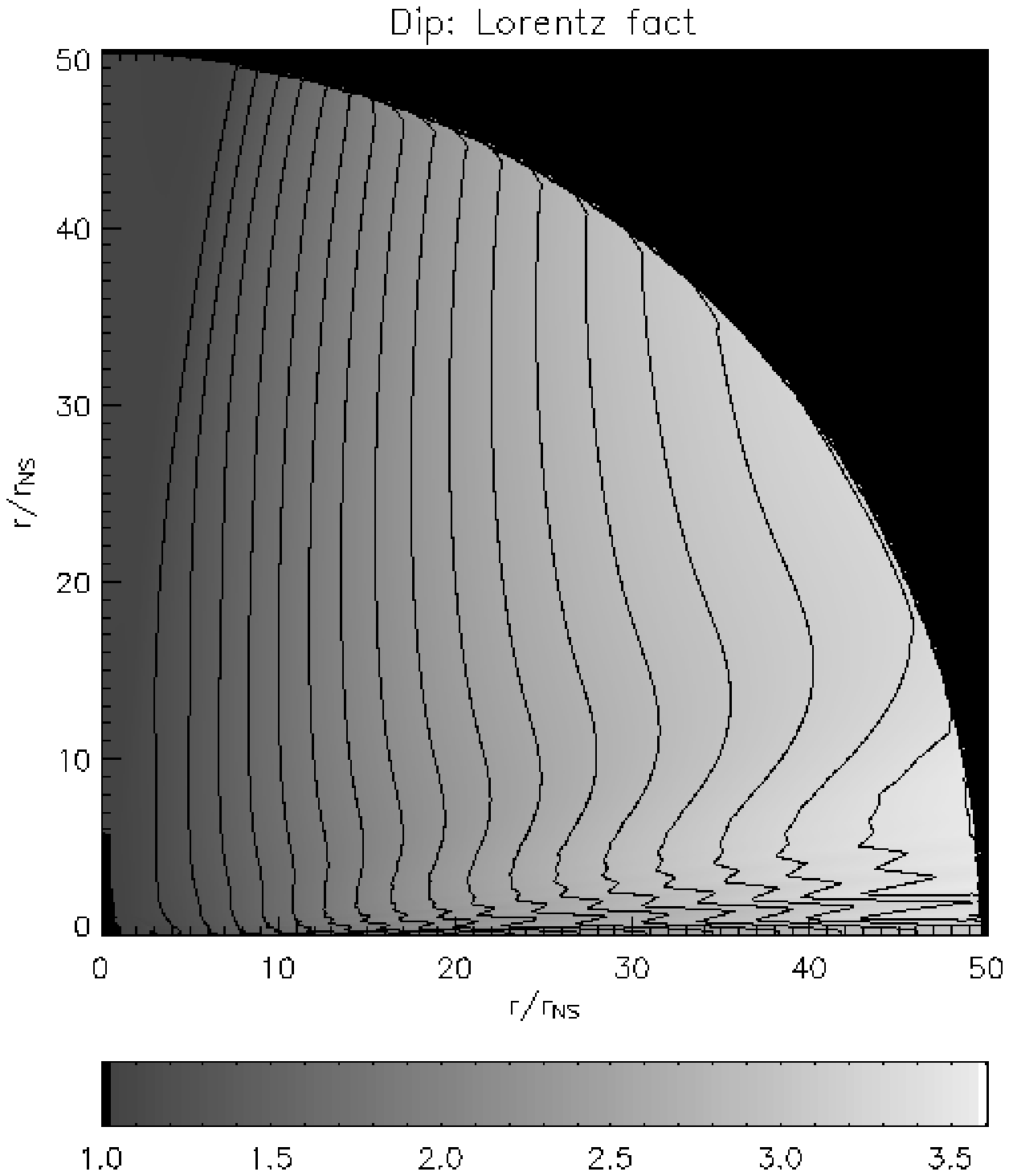,width=5cm}   }
 \caption{Wind structure for a dipolar magnetic field, $\sigma=20$, $\Omega=0.14$. 
 Left: poloidal magnetic field lines, and ratio $|B_{\phi}/B_r|$;  
 Alfvenic (dashed) and slow (dotted) surfaces. Center: same as left, note the monopolar 
 structure at larger radii. Fast (dotted) surface. Right: Lorentz factor, $\sim\sin{(\theta)}$. From \cite{buc06}.}
\end{figure*}

However, no analytic multidimensional solution of the full relativistic MHD 
equations exists that extends from the surface of the neutron star to large 
distances. Ambigiuties remain about injection conditions, which in the case 
of a newly-born hot magnetar, depend non-trivially on the temperature, gravity,
 and rotation. Even simplified models consider only supersonic injection, but 
none focus on the possibility of a thermal wind at the base.  Moreover the use of
 a monopolar field might prove not accurate, given than a more realistic dipole
 should be assumed in the close magnetophere.

\section{Wind acceleration}  

The first axisymmetric numerical simulations of the acceleration of a relativisitc
 wind were presented by \cite{bog99}, where the case of a zero pressure plasma 
in a monopolar magnetic field, with uniform injection conditions at the base 
of the star was investigated. Results show that as soon as $\gamma>5$ 
collimation and acceleration of the wind become marginal. 

Our recent investigation (Bucciantini et al. 2006) has focused on the study of the acceleration 
of an axisymmetric relativisitc wind in the case the pressure is 
small compared with the rest mass 
energy density, but not negligible. If acceleration is efficient, the wind can 
reaches high Lorentz factors ($\gamma\sim 100-300$) and kinetic luminosity 
($\sim 10^{50}$ erg s$^{-1}$), perhaps enough to trigger a GRB. The relativistic 
MHD equations, together with Maxwell equations, are solved using the
shock-capturing code for relativistic MHD developed by \cite{ldz03}. The code has 
been modified to solve the equations in the Schwarzschild metric, in order to 
properly account the effects of gravity. In order to model the heating and cooling
processes using an ideal gas equation of state we adopt an adiabatic
coefficient $\Gamma=1.1$ (almost isothermal wind). A hot atmosphere is assumed at 
the base of the star with $p/\rho c^2 \sim 0.01$. Density, pressure, rotation rate and 
the radial component of the magnetic field are fixed at the base of the neutron star,
 which is assumed to have a canonical radius $R_{NS}=10$ km. The system is let to
 relax to a steady state configuration. We verify that it takes less than 5 
rotations for the mass flux and energy losses to converge.

In analogy with the 1D model, solutions in the 2D case can be parametrized by 
using the average value of $\sigma$ over all stream lines. Results in the case 
of a monopolar magnetic field show that for $\sigma > 10$ the flow is close to 
radial, the terminal $\gamma$ is $< \sigma$, pointing to inefficient 
acceleration. Moreover the energy and angular momentum fluxes rapidly approach
 the force-free values. Even if $v << c$ inside the Light Cylinder, asymptotic
 results agree with the flow structure found by \cite{bog99}. Energy flux
 scales as $\sin^2{(\theta)}$, and  $\gamma\propto\sin{(\theta)}$, in agreement
 with the solution of the exact monopole. Mass flux is also shaped by rotation.
 On the equator it appears to be 5 times higher than in the non rotating case.
 On the contrary due to collimation in the polar region the mass loss rate 
along the axis is about twice smaller than in the non rotating case. Such 
variation is due to the change in centrifugal support at different latitudes,
 and decreases as the period increses. Results suggest that a rapidly rotating
 magnetar will not be able to simulataneusly produce a collimated and 
relativisitc outflow. Relativisitic velocities can be reached only for $\sigma>>1$, 
and in this case most of the energy would be concentrated in the equatorial plane
 instead that in a jet. Simulations have been extended to large distances
 ($\sim 10000 R_{NS}$) showing that for $\sigma>10$ most of the energy is 
in the equatorial plane. We observe the formation of a collimated narrow jet, 
that is unresolved in our grid, but such structure appears not to be energetically 
significant. 

\begin{figure}  
\centerline{\psfig{file=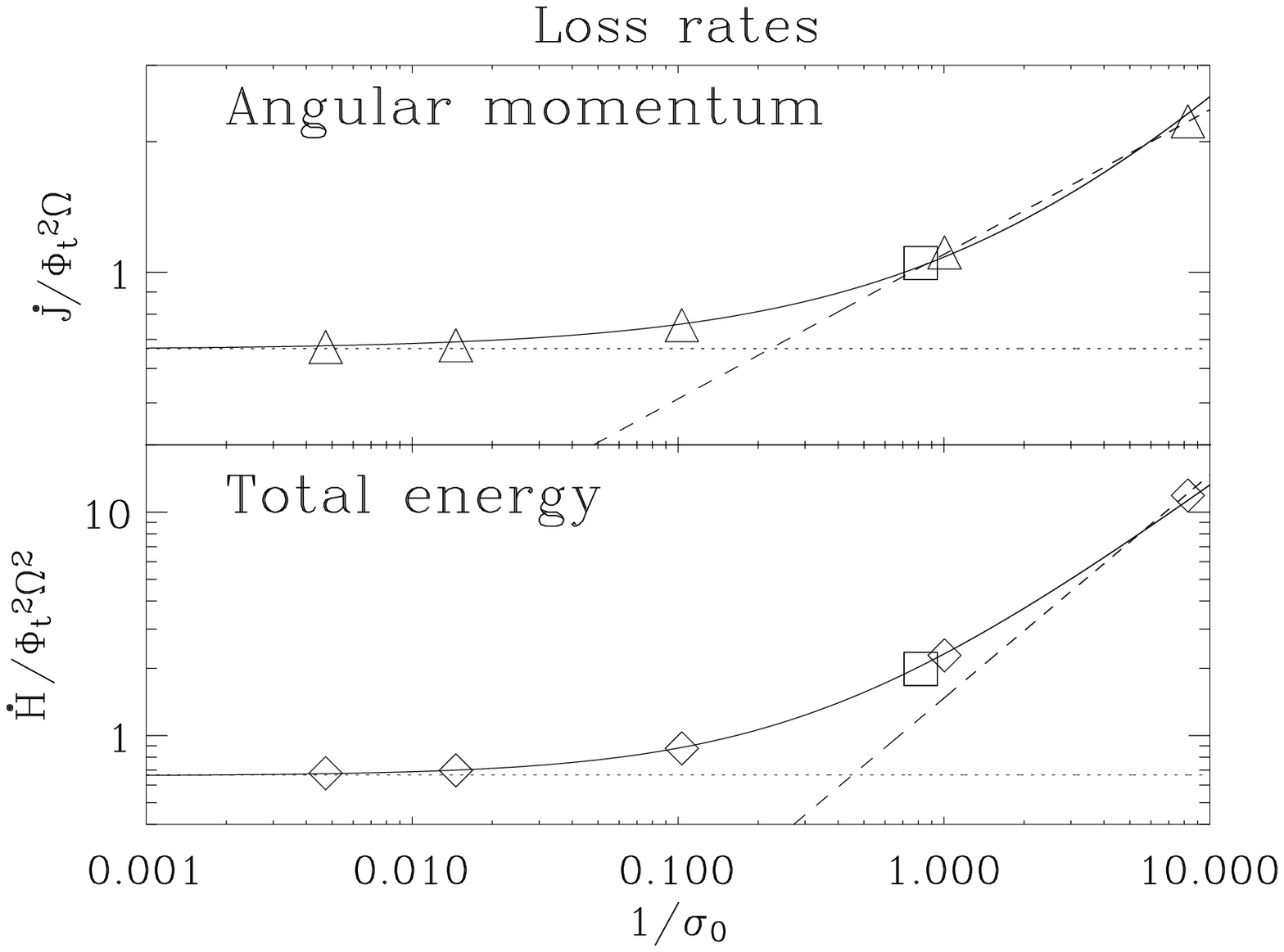,width=8cm} }
\centerline{\psfig{file=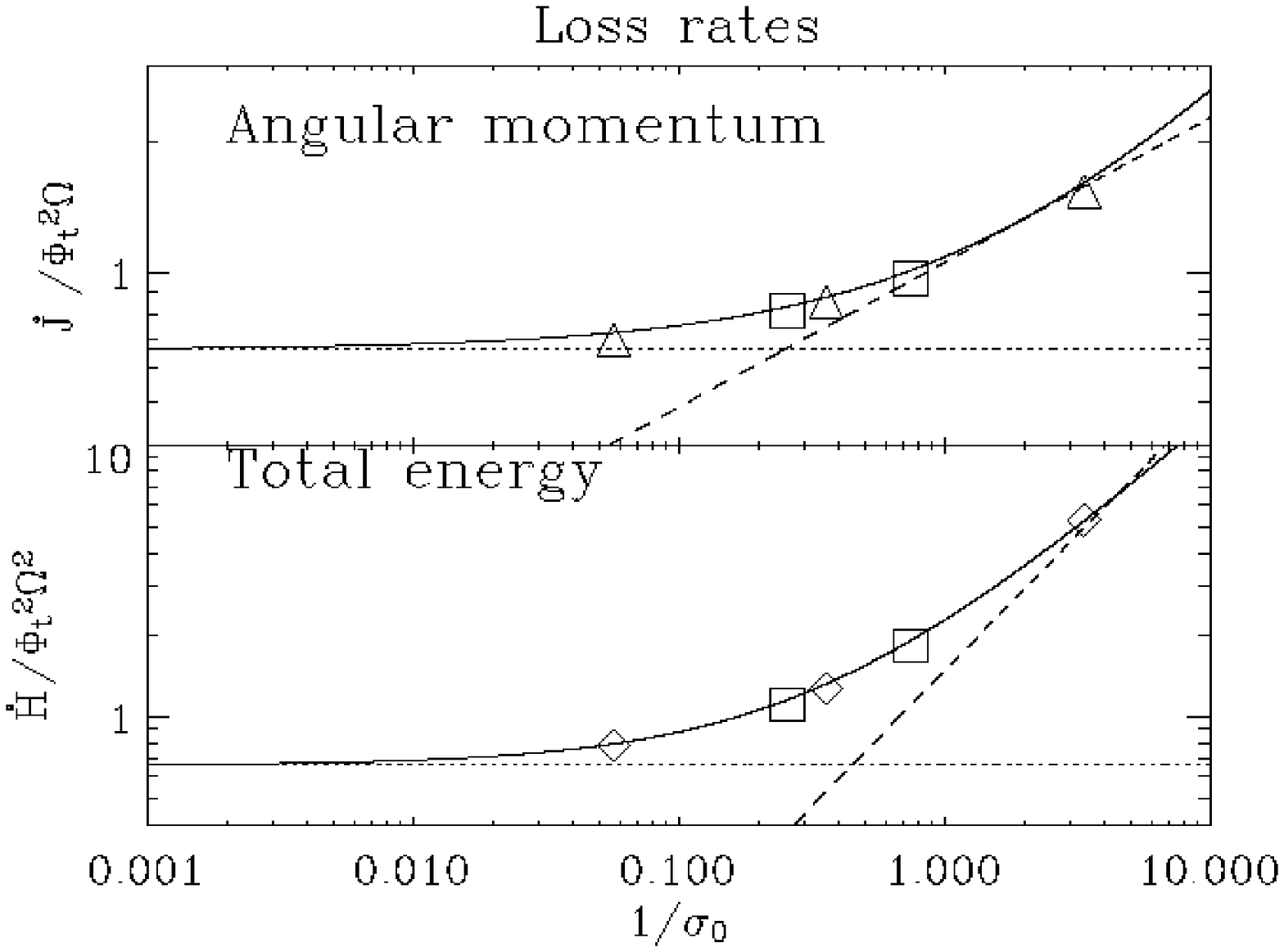,width=8cm} }
\caption{Energy and angular momentum losses for a monopolar (upper panel) and 
 dipolar (lower panel) magnetic field as a function of $\sigma_{o}$ ($\sigma$ 
 averaged on open field lines). Squares indicate cases with different rotation rate. 
 Dashed-lines represent the low-$\sigma$ analytic theory, dotted lines are the force-free limits. 
 The solid line is a power law fit for the monopolar field. When normalized on open field lines, 
 the results of the dipole follow the same curve of the monopole. From \cite{buc06}.}
\end{figure}

We computed solutions also in the more realistic cases of dipolar magnetic 
field at the surface of the NS. Due to stability issues in numerical RMHD it
 was not possible to study cases with $\sigma > 20$. In the monopolar case,
 no current sheet is present. 
In the case of a dipole  current sheets are present both at the edge of the 
closed zone and at the equator, and numerical resistivity is important.
 In particular in the equatorial plane, where
 the mangetic field vanishes, reconnection (due to numerical resistivity) 
takes place and plasmoids are formed. These plasmoids extend about 5-10 degrees 
around the equator, and cause fluctuation of about 10-20\% in the losses.
Even if such values depend on numerical resistivity, this suggests that reconnection
at the Y-point might manifest itself observationally. 

 To obtain steady solutions we suppressed
 resistivity at the equator, thus enhancing the instability of the code. 
Moreover, in the case of a dipole, the magnetic and flow surfaces are not 
aligned with the grid, and in this case inversion from conservative to
 primitive varibles is not stable for $B^2/\rho c^2 > 100$.    

Despite the presence of a closed zone, and the fact that within the Light 
Cylinder the field structure is close to a dipole, we found that solutions
 can again be parmaterized in term of $\sigma$, now defined as an average 
done only on open streamlines. In this case we find that energy and angular
 momentum losses scale as a monopole with the same $\sigma$.
 Again as soon as $\sigma> 1$ solutions for the energy and angular momentum losses
approach the force-free limit 
(\cite{cont99}), outside the Light Cylinder, and for $\sigma >10$ the 
asymptotic structure resembles the exact monopole, with energy flux peaking
 at the equator and a higher equatorial Lorentz factor. Similar results
 are found also by \cite{kom06}. We are able to related the amount of open 
magnetic flux to the distance of the Y-point from the surface of the star. 
Extrapolating to the force-free limits, our result is in agreement with \cite{gruz05}.

Contrary to expectations (\cite{cont99}) we also find that the location of 
the Y-point is inside the Light Cylinder. This is due to the presence of a 
finite pressure plasma at the base of the star. In this case the extent of the
 closed zone is limited by the pressure equilibrium at the Y-point. We also
verified that the ratio of the Y-point radius to the Light Cylinder $R_Y/R_{L}$
drops with decreasing $\Omega$, as one would expenct for a braking index less than 3.
 This implies that even if for large $\sigma$ the asymptotic wind structure is close to the 
force-free monopole solution, the spindown time can be shorter, and the 
total energy losses larger, that what is inferred by applying the dipole formula.

\section{Conclusions}

Recent numerical results have confirmed theoretical predictions about
 the acceleration of relativisitc winds from neutron stars. The conversion of magnetic
 to kinetic energy appears to be minimal, the flow closely resembles the 
expectation from the exact monopole solution, there is no evidence for 
the formation of a collimated energetic jet. The use of a more realistic
 dipolar field does not seem to change significanlty the wind properties
 far from the Light Cylinder. On the contrary the energy and angular
 momentum losses seem to depend only on the integrated value of $\sigma$,
 and, for $\sigma>1$, on the amount of open magnetic flux.

Regarding the possibility that a magnetar could trigger a GRB, we conclude that,
in ideal MHD, acceleration to relativistic speeds requires very high $\sigma$,
much higher than what is thought to characterized newly-born magnetars. Moreover 
there is no evidence of energy collimation at high magnetization. Even if
 dissipation of the current sheet could be invoked to accelerate the wind, 
it is not obvious that a jet will form, it is more likely that the 
energy will still be concentrated in the equatorial plane, as is suggested 
by X-ray images of Pulsar Wind Nebulae. 

Even if pulsar winds are not thermally driven at the base, they are cold, 
and the injection conditions depend on the physics of gaps, 
our results have several speculative implications for rotation powered
pulsars, where $\sigma$ is high. The fact that the Y-point is
 interior to the light cylinder (and the
Alfven radius), and that  $R_Y/R_L$  decreases as $\Omega$ decreases,
 suggests that observed braking indices less than 3
might be a consequence of pressure equilibrium. 
Even if globally, thermal pressure  is not 
likely to be relevant for classical pulsars,  magnetic dissipation at and 
near the $Y$ point might cause pressure forces to be significant in this localized
region.

The appearance of outwardly propagating plasmoids at and beyond the
Y-point as a consequence of (numerical) magnetic dissipation raises
the intriguing possibility that noise in pulsar spindown
 might arise from instability of the magnetospheric
currents. Torque fluctuations might give rise to noise and limit
cycles in the observed phase of pulsars' subpulses.
\vskip 0.4cm

\begin{acknowledgements}
We gratefully acknowledge the support by the WE-Heraeus Foundation. This work has 
been supported by NSF grant AST-0507813, and by NASA grant NAG5-12031. 
\end{acknowledgements}
   
